\documentclass[10pt]{iopart}

\usepackage{iopams}
\usepackage[T1]{fontenc}
\usepackage[utf8]{inputenc}
\usepackage{graphicx}
\usepackage{tabularx}
\usepackage{color, soul}
\usepackage{textcomp}
\usepackage{enumerate}
\usepackage{todonotes}
\usepackage[draft]{hyperref}
\usepackage{comment}
\usepackage{subcaption}
\usepackage{url}
\usepackage[normalem]{ulem}
\usepackage{makecell}

\renewcommand{\vec}[1]{\mathbf{#1}}

\providecommand{\xl}[1]{#1}

\begin{document}

\title[A computational study of steady and stagnating positive streamers in N$_2$-O$_2$ mixtures]{A computational study of steady and stagnating positive streamers in N$_2$-O$_2$ mixtures}

\author{Xiaoran Li$^{1,2}$, Baohong Guo$^{1}$, Anbang Sun$^{2}$, Ute Ebert$^{1,3}$, Jannis Teunissen$^1$}

\address{$^1$Centrum Wiskunde \& Informatica, Amsterdam, The Netherlands\\
        $^2$State Key Laboratory of Electrical Insulation and Power Equipment, School of Electrical Engineering,
        Xi'an Jiaotong University, Xi'an, 710049, China\\
        $^3$ Eindhoven University of Technology, Eindhoven, The Netherlands}
\ead{jannis.teunissen@cwi.nl}
\vspace{10pt}
\begin{indented}
        \item[]\today
\end{indented}

\begin{abstract}
  In this paper, we address two main topics: steady propagation fields for positive streamers in air and streamer deceleration in fields below the steady propagation field.
  We generate constant-velocity positive streamers in air with an axisymmetric fluid model, by \xl{initially} adjusting the applied voltage based on the streamer velocity.
  After an initial transient, we observe steady propagation for velocities of $3\times10^4$\,m/s to $1.2\times10^5$\,m/s, during which streamer properties and the background field do not change.
  This propagation mode is not fully stable, in the sense that a small change in streamer properties or background field \xl{eventually} leads to acceleration or deceleration.
  An important finding is that faster streamers are able to propagate in significantly lower background fields than slower ones, indicating that there is no unique stability field.
  We relate the streamer radius, velocity, maximal electric field and background electric field to a characteristic time scale for the loss of conductivity.
  This relation is qualitatively confirmed by studying streamers \xl{in N$_2$-O$_2$ mixtures with less oxygen than air.}
  In such mixtures, steady streamers require lower background fields, due to a reduction in the attachment and recombination rates.
  We also study the deceleration of streamers, which is important to predict how far they can propagate in a low field.
  Stagnating streamers are simulated by applying a constant applied voltage.
  We show how the properties of these streamers relate to the steady cases, and present a phenomenological model with fitted coefficients that describes the evolution of the velocity and radius.
  Finally, we compare the lengths of the stagnated streamers with predictions based on the conventional stability field.
\end{abstract}

\ioptwocol

\section{Introduction}
\label{sec:introduction}


Streamer discharges~\cite{nijdam2020, ebert2010} are a common initial stage of electrical
discharges, playing an important role for electric breakdown in nature and in high voltage technology.
As a cold atmospheric plasma (CAP)~\cite{weltmann2019} they also have wide industrial applications~\cite{bardos2010,laroussi2014,popov2016}.
The goal of this paper is to better understand positive streamer propagation in air.
In particular, we study when such streamers accelerate or decelerate in homogeneous fields, by locating the unstable boundary between these regimes with numerical simulations.

An important empirical concept for streamer propagation has been the ``stability field'' $E_\mathrm{st}$~\cite{gallimberti1979},
which is often defined as the minimal background electric field that can sustain streamer propagation.
Experimentally, stability fields have been extensively investigated~\cite{phelps1971, griffiths1976, allen1991, allen1995,veldhuizen2002, seeger2018}.
For positive streamers in air with standard humidity (11\,g/m$^3$), reported values range from 4.1 to 6\,kV/cm, with values around 5\,kV/cm being the most common.
Note that if there are multiple streamers, they will modify the background field in which each of them propagates.
However, the small spread in experimental measurements indicates that the concept of a stability field nevertheless remains useful.

In~\cite{gallimberti1979, phelps1971} it was suggested that a streamer would propagate with a constant velocity and radius in the stability field.
This led to the concept of a ``steady propagation field'' in which streamer properties like velocity and radius do not change~\cite{qin2014, francisco2021d}.
Such steady propagation was recently observed in numerical simulations in air in a field of about 4.7\,kV/cm~\cite{francisco2021d}.
\xl{In these simulations, the conductivity behind the streamer head was lost after a certain length due to electron attachment and recombination.}
The resulting discharge resembled the minimal streamers found in~\cite{briels2008a}.
Qin \textit{et al.}~\cite{qin2014} proposed that steady propagation fields in air depend on streamer properties and that they could be as high as the breakdown electric field (28.7\,kV/cm), based on energy conservation criteria~\cite{raizer1991}.

In this paper, we address two main topics. The first is to study steady propagation fields for positive streamers in air, in particular the range of such fields and their dependence on streamer properties.
The second topic is how streamers decelerate in fields lower than their steady propagation field, and whether the lengths of such streamers can be predicted.

Below, we briefly summarize some of the past work on decelerating and stagnating streamers.
Pancheshnyi \textit{et al.}~\cite{pancheshnyi2004} numerically investigated the stagnation dynamics of positive streamers with an axisymmetric fluid model.
It was shown that the streamer's radius decreased as it decelerated, which led to a rapid increase in the electric field at the streamer head.
More recently, Starikovskiy \textit{et al.}~\cite{starikovskiy2021} studied decelerating streamers in an inhomogeneous gas density with an axisymmetric fluid model.
Among other things, the authors demonstrated the rather different stagnation dynamics of positive and negative streamers.
In~\cite{francisco2021d}, it was observed that positive streamers decelerate and eventually stagnate in a background electric field below their steady propagation field.
With a standard fluid model with the local field approximation,
the electric field at the streamer head diverges as the streamer stagnates.
In~\cite{niknezhad2021}, suitable models for simulating positive streamer stagnation were investigated, and it was shown that the field divergence can be avoided by using an extended fluid model.
In this paper, we instead modify the impact ionization source term to avoid this unphysical divergence~\cite{soloviev2009a,teunissen2020}.

In this paper, we also study the relation between the properties of steady streamers.
Several relations have been found in past studies, in particular between the streamer velocity and radius.
In the experimental work of Briels \textit{et al}~\cite{briels2008a}, the velocity $v$ \xl{was parameterized} in terms of the diameter $d$ as $v = 0.5 d^2 \textrm{mm}^{-1} \textrm{ns}^{-1}$, for both positive and negative streamers in an inhomogeneous field.
In contrast, simulation results in~\cite{luque2008a} indicated an approximately linear relation between streamer velocity and radius for accelerating streamers.
Approximate analytic results in~\cite{naidis2009} supported this quasi-linear relation, and it was shown that with certain assumptions, the maximum electric field at the streamer head can be determined from $v$ and $d$.

We simulate the propagation of positive streamers in air with a 2D axisymmetric fluid model, which is described in section \ref{sec:conditions}.
In section \ref{sec:invest-steady-stre}, we investigate the properties of steady streamers, which are obtained by adjusting the applied voltage based on the streamer velocity.
Afterwards, the deceleration of streamers is studied, by simulating stagnating streamers in a low background field in section \ref{sec:stopping}.



\section{Simulation model}
\label{sec:conditions}

\subsection{Fluid model and chemical reactions}
We use a 2D axisymmetric drift-diffusion-reaction type fluid model with the
local field approximation, as implemented in the open-source \texttt{Afivo-streamer}~\cite{teunissen2017} code.
For a recent comparison of experiments and simulations using \texttt{Afivo-streamer} see~\cite{xiaoran-comparison},
and for a comparison between \texttt{Afivo-streamer} and five other simulation codes see~\cite{bagheri2018}.
Furthermore, in~\cite{Wang_2022}, simulations with fluid model used here were compared against particle-in-cell simulations in 2D and 3D, generally finding good agreement.

In the model, both electron and ion densities evolve due to transport and reaction terms.
The temporal evolution of the electron density ($n_e$) is given by
\begin{equation}
    \partial_t n_e = \nabla \cdot (n_e \mu_e \vec{E} + D_e \nabla n_e) + S,
\end{equation}
where $\mu_e$ is the electron mobility, $\vec{E}$ the electric field, $D_e$ the
electron diffusion coefficient and $S$ is the sum of source terms, given by
\begin{equation}
  \label{eq:sources}
  S = S_i - S_\eta + S_\mathrm{detach} - S_\mathrm{recom} + S_\mathrm{ph},
\end{equation}
where $S_i$, $S_\eta$, $S_\mathrm{detach}$, $S_\mathrm{recom}$ and $S_\mathrm{ph}$ are the source terms for impact ionization, attachment, detachment, electron-ion recombination and non-local photoionization, respectively.
Photoionization is computed according to Zheleznyak's model~\cite{zheleznyak1982} using the Helmholtz approximation~\cite{luque2007,bourdon2007},
using the same photoionization model as~\cite{francisco2021d}.

The chemical reactions considered in this paper are listed in table \ref{tbl:reaction_table}.
They include electron impact
ionization ($k_1$–$k_3$), electron attachment ($k_4$, $k_5$), electron detachment ($k_6$-$k_8$), ion conversion ($k_9$-$k_{13}$) and electron-ion recombination ($k_{14}$, $k_{15}$).
The electron transport data and the electron impact reaction coefficients depend on
the reduced electric field $E/N$, and they were computed using
BOLSIG+ \cite{hagelaar2005} with Phelps’ cross sections for (N$_2$, O$_2$)
\cite{phelps1985, Phelps-database} using a temporal growth model.
As was pointed out in~\cite{Wang_2022}, data computed with a temporal growth model is more suitable for positive streamer simulations than data computed with a spatial growth model, which was used in \cite{francisco2021d}.


Our model includes ion motion, which can be important at relatively low streamer velocities or when studying streamer stagnation~\cite{niknezhad2021}.
The temporal evolution of each of the ion species $n_j$ listed in table \ref{tbl:reaction_table} is described by
\begin{equation}
  \partial_t n_j = -\nabla \cdot (\pm n_j \mu_\mathrm{ion} \vec{E}) + S_j,
\end{equation}
where $S_j$ is the sum of source terms for these species, and $\mu_\mathrm{ion}$ is the ion mobility, $\pm$ is the sign of the species’ charge.
For simplicity, we use a constant ion mobility $\mu_\mathrm{ion} = 2.2 \times 10^{-4}$\,m$^2$/Vs~\cite{tochikubo2002} for all ion species, as was also done in~\cite{francisco2021d}.

The electric field $\vec{E}$ is computed as $\vec{E}=-\nabla \varphi$ after solving Poisson's equation
\begin{equation}
\nabla \cdot \left(\varepsilon_0 \nabla \varphi \right)=-\rho,
\end{equation}
where $\varepsilon_0$ is the vacuum permittivity and $\rho$ is the space charge density.

It can be difficult to simulate slow or stagnating positive streamers with a standard fluid model using the local field approximation~\cite{niknezhad2021}.
Such streamers have a small radius, leading to strong electron density and field gradients that reduce the validity of the local field approximation~\cite{naidis1997, li2010a}.
In~\cite{francisco2021d, pancheshnyi2004, starikovskiy2021} the electric field at the streamer tip was found to rapidly increase during streamer stagnation, and simulations had to be stopped after the field became unphysically large.
Recently, it was shown that such unphysical behavior can be avoided by using an extended fluid model that
includes a source term correction depending on $\nabla n_e$~\cite{niknezhad2021}.
In this paper, we instead use a correction factor $f_\epsilon$ for the impact ionization term as described in~\cite{soloviev2009a,teunissen2020}.
This factor is given by
\begin{equation}
  \label{eq:factor}
  f_\epsilon = 1 - \frac{\hat{\vec{E}} \cdot \vec{\Gamma}^{\mathrm{diff}}}{||\vec{\Gamma}^{\mathrm{drift}}||},
\end{equation}
where $\hat{\vec{E}}$ is the electric field unit vector, and $\vec{\Gamma}^{\mathrm{diff}}$ and $\vec{\Gamma}^{\mathrm{drift}}$ are the diffusive and drift flux of electrons, respectively,
and $f_\epsilon$ is limited to the range of $[0,1]$.
As discussed in~\cite{soloviev2009a,teunissen2020}, this correction factor prevents unphysical growth of the plasma near strong density and field gradients.
The underlying idea is that the diffusive electron flux parallel to the electric field (thus corresponding to a loss of energy) should not contribute to impact ionization.

\begin{table*}[ht]
        \centering
        \caption{Reactions included in the model, with reaction rates and
                references. The electron temperature $T_e$ in reaction rates $k_{14}$ and $k_{15}$ is obtained from the mean electron energy ($\epsilon_e$) computed by BOLSIG+ as \mbox{$T_e=2\epsilon_e/3k_B$}.}
        \begin{tabular}{cccc}
                \hline
                Reaction No. & Reaction & Reaction rate coefficient  & Reference \\ \hline
                1 & $\textrm{e} + \textrm{N}_2 \stackrel{k_1}{\longrightarrow} \textrm{e} + \textrm{e} + \textrm{N}_2^+$ (15.60 eV) & $k_1(E/N)$ &~\cite{phelps1985, hagelaar2005} \\
                2 & $\textrm{e} + \textrm{N}_2 \stackrel{k_2}{\longrightarrow} \textrm{e} + \textrm{e} + \textrm{N}_2^+$ (18.80 eV) & $k_2(E/N)$ &~\cite{phelps1985, hagelaar2005} \\
                3 & $\textrm{e} + \textrm{O}_2 \stackrel{k_3}{\longrightarrow} \textrm{e} + \textrm{e} + \textrm{O}_2^+$ & $k_3(E/N)$ &~\cite{phelps1985, hagelaar2005} \\
                4 & $\textrm{e} + \textrm{O}_2 + \textrm{O}_2 \stackrel{k_4}{\longrightarrow} \textrm{O}_2^- + \textrm{O}_2$    & $k_4(E/N)$ &~\cite{phelps1985, hagelaar2005}  \\
                5 & $\textrm{e} + \textrm{O}_2 \stackrel{k_5}{\longrightarrow} \textrm{O}^- + \textrm{O}$                       & $k_5(E/N)$ &~\cite{phelps1985, hagelaar2005}  \\
                6 & O$_2^-$ + N$_2$ $\stackrel{k_6}{\longrightarrow}$ O$_2$ + N$_2$ + e &
                $k_6=1.13\times10^{-25}\textrm{m}^3\textrm{s}^{-1}$ & \cite{kossyi1992}  \\
                7 & O$_2^-$ + O$_2$ $\stackrel{k_7}{\longrightarrow}$ O$_2$ + O$_2$ + e  &
                $k_7=2.2\times10^{-24}\textrm{m}^3\textrm{s}^{-1}$ & \cite{kossyi1992} \\
                8 & $\textrm{O}^- + \textrm{N}_2 \stackrel{k_{8}}{\longrightarrow} \textrm{e} + \textrm{N}_2\textrm{O}$ & $k_{8}=1.16\times10^{-18}\exp(-(\frac{48.9}{11+E/N})^2)\,\textrm{m}^3\textrm{s}^{-1}$ & \cite{pancheshnyi2013} \\
                9 & $\textrm{O}^- + \textrm{O}_2 \stackrel{k_{9}}{\longrightarrow} \textrm{O}_2^- + \textrm{O}$ & $k_{9}=6.96\times10^{-17}\exp(-(\frac{198}{5.6+E/N})^2)\,\textrm{m}^3\textrm{s}^{-1}$ & \cite{pancheshnyi2013} \\
                10 & $\textrm{O}^- + \textrm{O}_2 + \textrm{M} \stackrel{k_{10}}{\longrightarrow} \textrm{O}_3^- + \textrm{M}$ & $k_{10}=1.1\times10^{-42}\exp(-(\frac{E/N}{65})^2)\,\textrm{m}^6\textrm{s}^{-1}$ & \cite{pancheshnyi2013} \\
                11 & $\textrm{N}_2^+ + \textrm{N}_2 + \textrm{M} \stackrel{k_{11}}{\longrightarrow} \textrm{N}_4^+ + \textrm{M}$ & $k_{11}=5\times10^{-41}\,\textrm{m}^6\textrm{s}^{-1}$ & \cite{aleksandrov1999} \\
                12 & $\textrm{N}_4^+ + \textrm{O}_2 \stackrel{k_{12}}{\longrightarrow} \textrm{O}_2^+ + \textrm{N}_2 + \textrm{N}_2$ & $k_{12}=2.5\times10^{-16}\,\textrm{m}^3\textrm{s}^{-1}$ & \cite{aleksandrov1999} \\
                13 & $\textrm{O}_2^+ + \textrm{O}_2 + \textrm{M} \stackrel{k_{13}}{\longrightarrow} \textrm{O}_4^+ + \textrm{M}$ & $k_{13}=2.4\times10^{-42}\,\textrm{m}^6\textrm{s}^{-1}$ & \cite{aleksandrov1999} \\
                14 & $\textrm{e} + \textrm{O}_4^+ \stackrel{k_{14}}{\longrightarrow} \textrm{O}_2 + \textrm{O}_2$     & $k_{14}(E/N)=1.4\times10^{-12}(300\textrm{K}/T_e)^{1/2}\,\textrm{m}^3\textrm{s}^{-1}$ & \cite{kossyi1992} \\
                15 & e + N$_4^+$ $\stackrel{k_{15}}{\longrightarrow}$ N$_2$ + N$_2$ & $k_{15}(E/N)=2.0\times10^{-12}(300\textrm{K}/T_e)^{1/2}\,\textrm{m}^3\textrm{s}^{-1}$ & \cite{kossyi1992} \\
                \hline
        \end{tabular}
        \label{tbl:reaction_table}
\end{table*}

\subsection{Computational domain and initial conditions}

\begin{figure}
        \centering
        \includegraphics[width=0.5\textwidth]{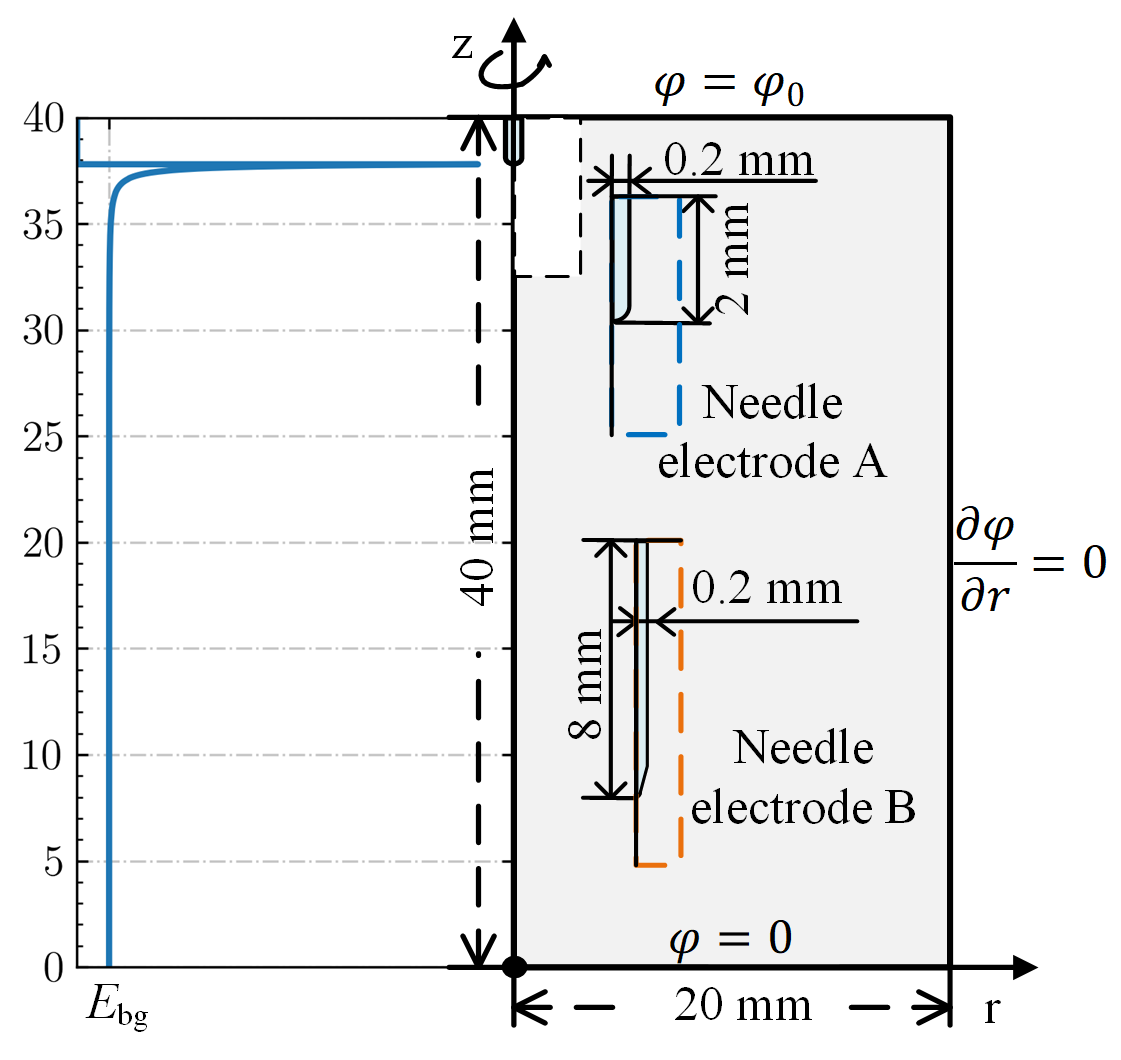}
        \caption{The computational domain. Right: electrode geometry and boundary conditions. The needle protrusions used for generating constant-velocity (needle electrode A) and stagnating streamers (needle electrode B) are illustrated. Left: axial electric field distribution with the 2 mm long needle. $E_\mathrm{bg}$ is the average electric field between the plate electrodes.}
        \label{fig:computational_domain}
\end{figure}

We simulate positive streamers in N$_2$-O$_2$ mixtures at 300 K and 1 bar, using the axisymmetric computational domain illustrated in figure \ref{fig:computational_domain}.
A high voltage is applied to the upper plate electrode (at $z=40$\,mm), which includes a needle protrusion.
For the constant-velocity streamers simulated in section \ref{sec:invest-steady-stre}, this needle is 2\,mm long, with a radius of 0.2\,mm and a semispherical tip.
To generate the stagnating streamers simulated in section \ref{sec:stopping}, more field enhancement is required.
The needle used there is 8\,mm long, with a radius of 0.2\,mm, a conical tip with 60$^\circ$ top angle and a tip curvature radius of 50\,$\mu$m.
The lower plate electrode (at $z=0$\,mm) is grounded.
A homogeneous Neumann boundary condition is applied for the electric potential
on the radial boundary.
For plasma species densities, homogeneous Neumann boundary conditions are used on all the domain boundaries.
However, at the positive high-voltage electrode electron fluxes are absorbed but not emitted.

With the short needle electrode the axial electric field is approximately uniform, except for a small area around the needle tip, as shown in figure~\ref{fig:computational_domain}.
For $z$ between 0\,mm and 32\,mm, the electric field differs less than 1\% from the average electric field between the plates $E_\mathrm{bg}$.
We therefore refer to $E_\mathrm{bg}$ as ``the background electric field'' in the rest of the paper.

To initiate the discharges, a neutral Gaussian seed consisting of electrons and positive ions (N$_2^+$) is used.
Its density is given by $n_0$ exp$(-(d/R)^2)$, with $n_0$ = $10^{14}$ m$^{-3}$, $d$
the distance to the needle tip and $R$ = 5\,mm.
Besides this initial seed, no other initial ionization is included.

Adaptive mesh refinement is used in the model for computational efficiency.
The refinement criterion for the grid spacing $\Delta x$ is based on $1/\alpha(E)$, where $\alpha (E)$ is the field-dependent ionization coefficient.
If $\Delta x > c_0 / \alpha (E)$, the mesh is refined, and if $\Delta x < \min\{0.125 \, c_0 / \alpha (E),\, d_0\}$, the mesh is de-refined.
We use $c_0 = 0.5$ and $d_0 = 10\,\mu$m, which leads to a minimal grid spacing of $\Delta x_\mathrm{min} = 1.4\,\mu$m.

\subsection{Velocity control method}
\label{sec:velocity_control}

In this paper, we study steady streamers, propagating at a constant velocity.
To generate such streamers, we adjust the applied voltage $\phi$ based on the difference between the present streamer velocity $v$ and a goal velocity $v_{\mathrm{goal}}$.
In the simulations, we cannot accurately measure $v$ at every time step.
Instead, we take samples $v^*$ after the streamer head has moved more than $8 \, \Delta x_\mathrm{min}$
$$v^* = \left(z_\mathrm{head}^i - z_\mathrm{head}^{i-1}\right)/\left(t^i-t^{i-1}\right),$$
where $z_\mathrm{head}$ is the location of the maximum electric field and the superscripts $^i$ and $^{i-1}$ indicate the present and previous sampling time.
Since this estimate is still rather noisy, we average it with the four most recent samples of $v$ to obtain a smoothed velocity $v^i$. The voltage is then updated as
\begin{equation}
  \label{eq:velcontrol}
  \phi^i = \phi^{i-1} +  K_p (t^i-t^{i-1}) (1 - v^i/v_{\mathrm{goal}}),
\end{equation}
where $K_p$ is a proportionality constant between $5\times10^{13}$\,V/s and $2\times10^{14}$\,V/s.

\begin{figure}
        \centering
        \includegraphics[width=0.5\textwidth]{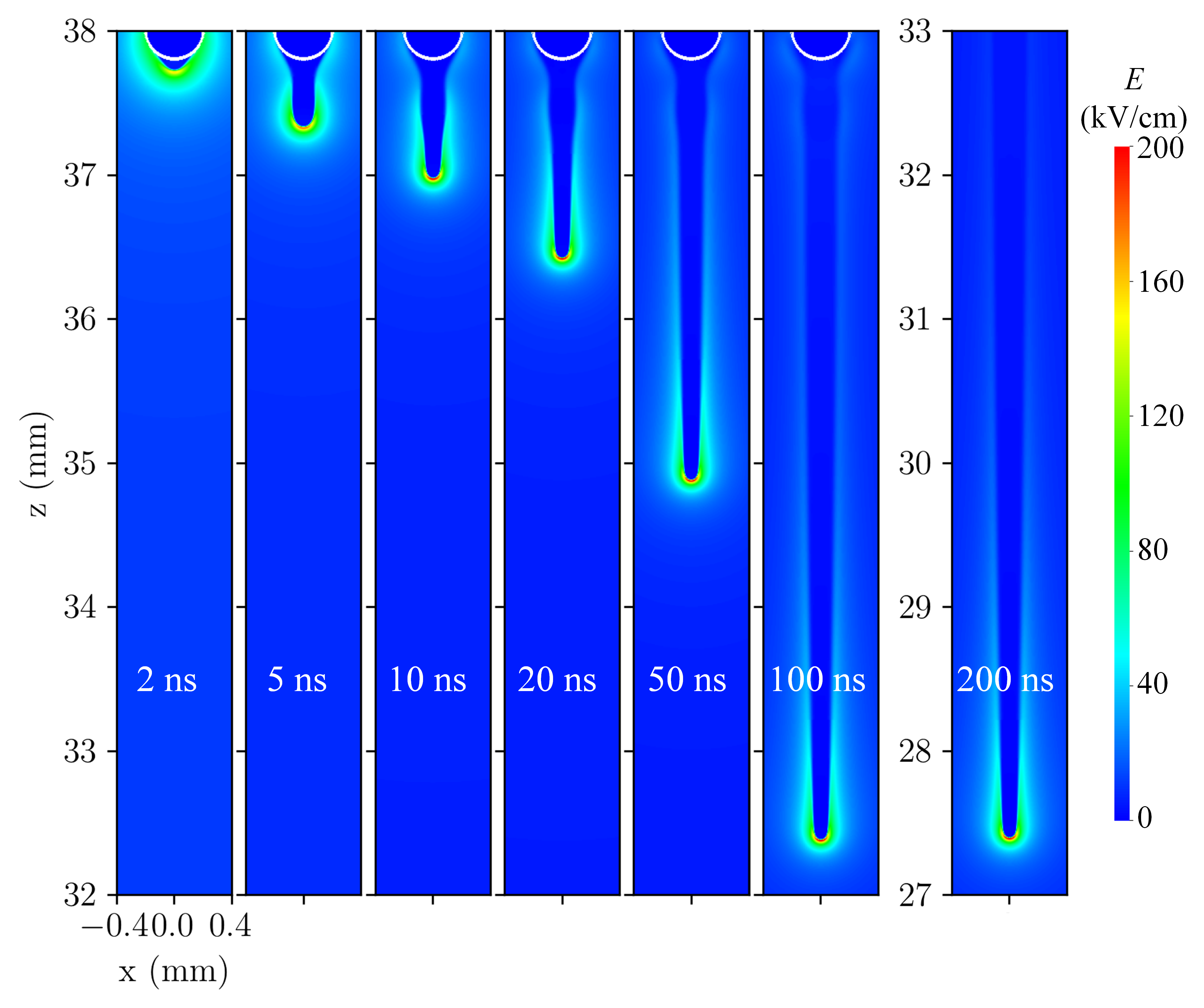}
        \caption{Electric field strength between 2\,ns and 200\,ns for a streamer whose velocity is forced to $5\times10^4$\,m/s by the velocity control method described in section~\ref{sec:velocity_control}.
        Note that the rightmost sub-figure has a different z axis.
        The white curves indicate the tip of the needle electrode.}
        \label{fig:method_2d}
\end{figure}

\begin{figure}
        \centering
        \includegraphics[width=0.5\textwidth]{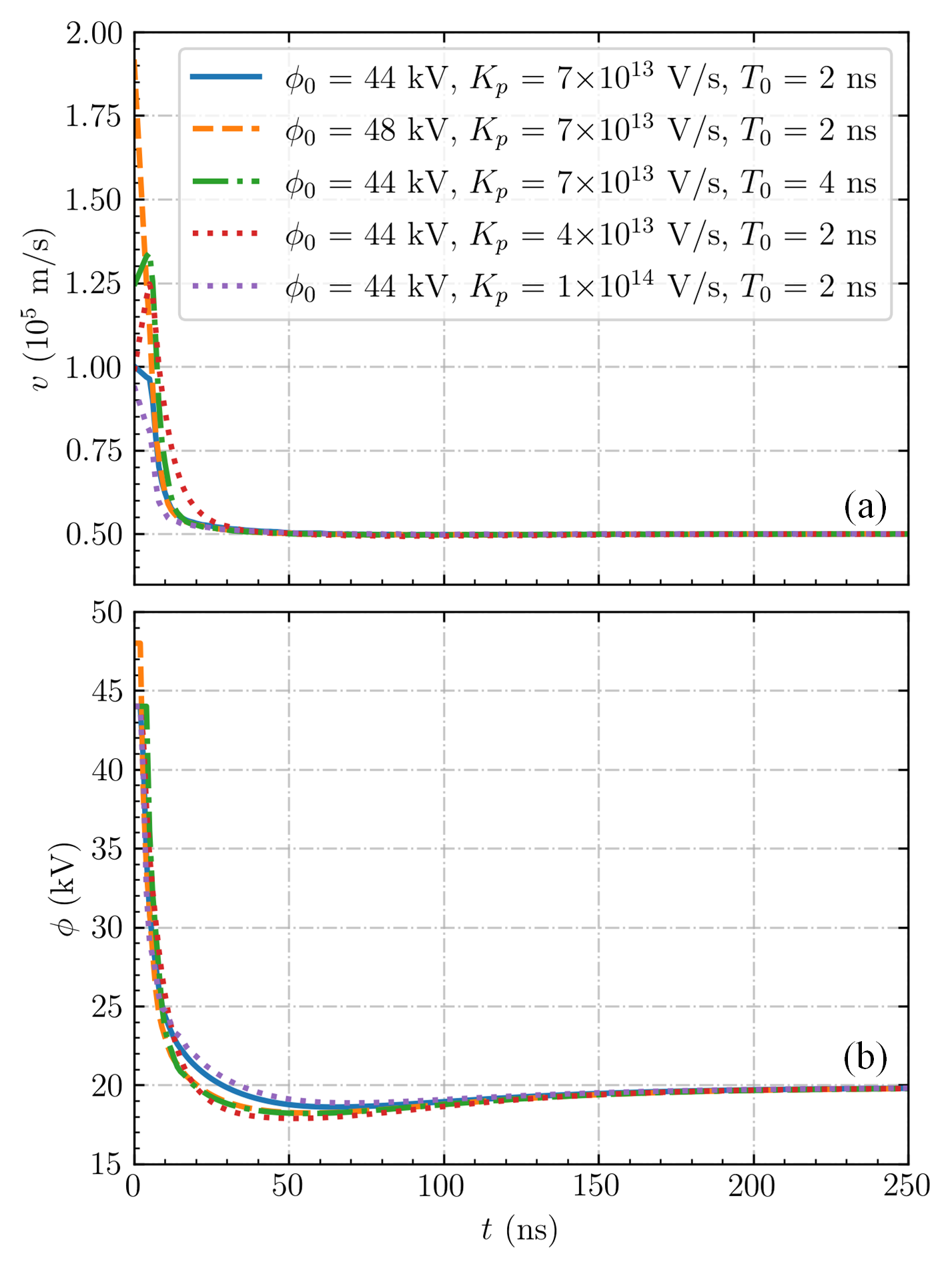}
        \caption{The streamer velocity (a) and applied voltage (b) versus time for streamers whose velocities are forced to be $5\times10^4$\,m/s but with different initially applied voltages $\phi_0$, the start time of voltage adjustment $T_0$ and the proportional coefficients $K_p$ in the velocity control method.}
        \label{fig:velocity_cotrol}
\end{figure}

Figure~\ref{fig:method_2d} shows an example of a streamer forced to propagate at $5\times10^4$\,m/s.
Corresponding profiles for the streamer velocity and applied voltage are shown in figure~\ref{fig:velocity_cotrol} (the solid line).
Initially, the applied voltage is 44\,kV, which corresponds to a background electric field of 11\,kV/cm.
The applied voltage is adjusted after 2\,ns, using $K_p$ $7\times10^{13}$\,V/s.
As the applied voltage is reduced, the streamer velocity decreases, until it eventually converges to the goal velocity after about 50\,ns.
The applied voltage then slightly increases, until it stabilizes after about 200\,ns.

The initially applied voltage ($\phi_0$), the delay until the voltage is first adjusted ($T_0$) and the coefficient $K_p$ can affect how the streamer velocity approaches the goal value.
Figure~\ref{fig:velocity_cotrol} shows the streamer velocities (a) and applied voltages (b) versus time for streamers whose velocities are forced to be $5\times10^4$\,m/s but with different $\phi_0$, $T_0$ and $K_p$.
Although the initial profiles vary, they eventually converge to the same value, which is also true for the streamer radius and the maximal electric field (which are not shown here).
This example therefore indicates that there is a unique propagation mode for a given constant streamer velocity.
In the rest of the paper we will use $\phi_0$ = 44\,kV, $T_0$ = 2\,ns, and $K_p$ = $7\times10^{13}$\,V/s, unless stated otherwise.



\section{Investigation of steady streamers}
\label{sec:invest-steady-stre}

In this section, we investigate ``steady streamers'' at constant velocities.
We remark that these streamers are not actually stable, in the sense that a small change in their properties would lead to either acceleration or deceleration. 
The streamers studied here thus demarcate the unstable boundary between acceleration and deceleration.

\subsection{Steady propagation in different background electric fields}
\label{sec:diffv}

\begin{figure*}[th]
        \centering
        \includegraphics[width=0.85\linewidth]{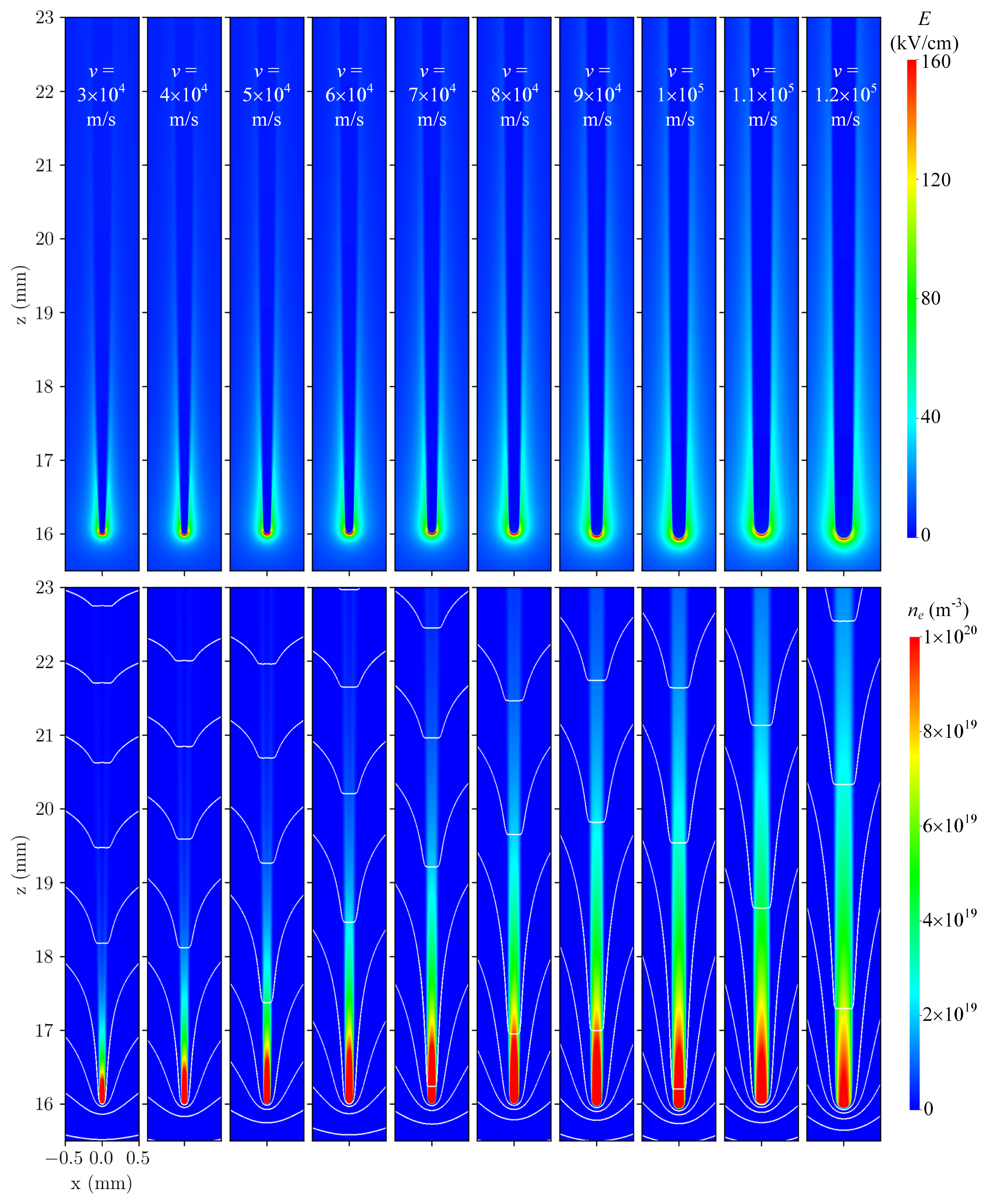}
        \caption{The electric field strength and the electron density for streamers in air at velocities of $3\times10^4$\,m/s to $1.2\times10^5$\,m/s, when their heads are at z = 16\,mm.
          All these streamers have reached steady states.
          The white equipotential lines are spaced by 0.5\,kV.
          Behind the streamer heads, the radius increases due to ion motion.}
        \label{fig:diffv_2d}
      \end{figure*}

\begin{figure}
        \centering
        \includegraphics[width=0.5\textwidth]{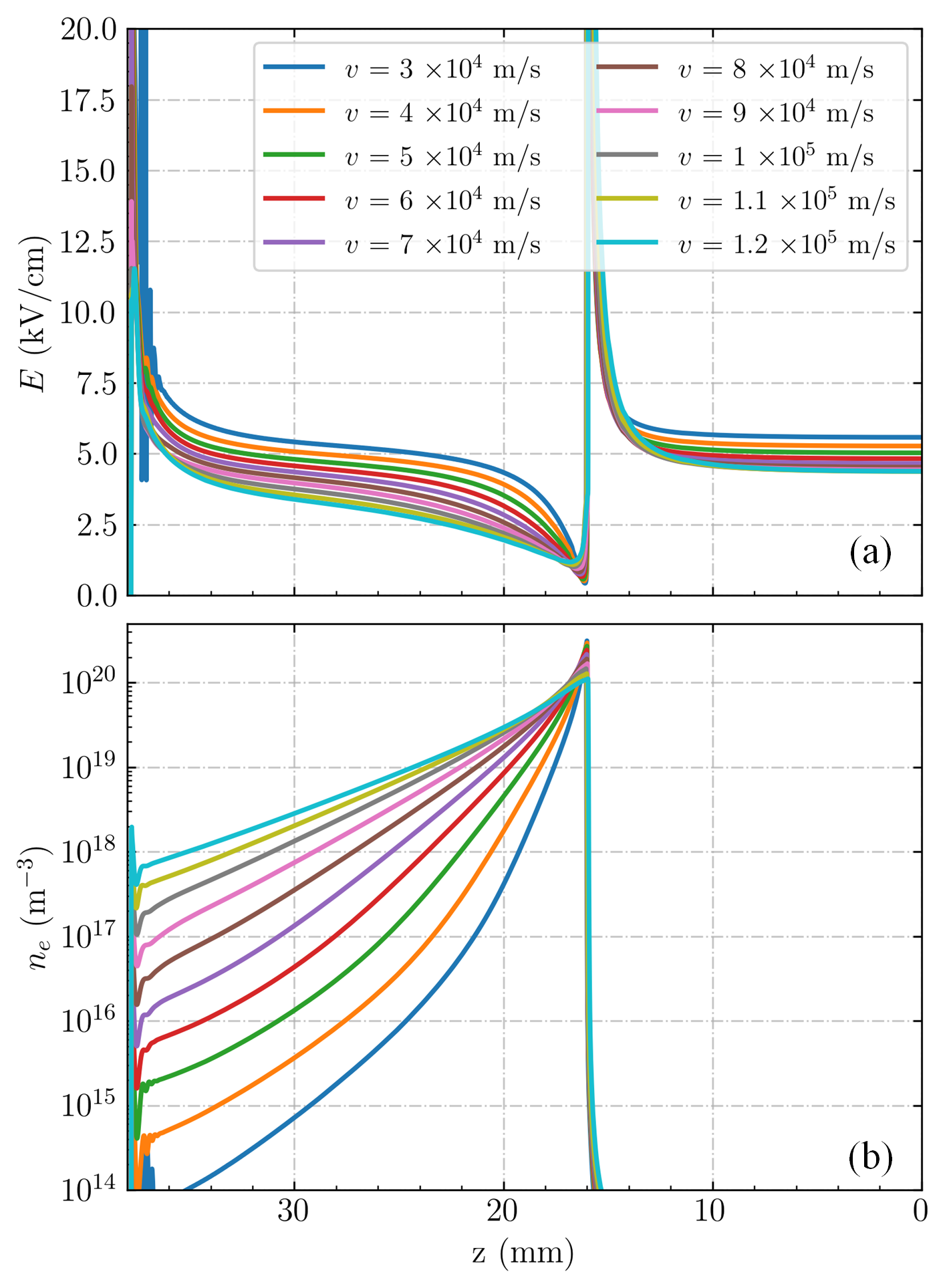}
        \caption{The axial electric field strength (a) and electron density (b)
          corresponding to figure~\ref{fig:diffv_2d}.}
        \label{fig:onaxis_infor}
\end{figure}

\begin{figure*}[th]
        \centering
        \includegraphics[width=1.0\linewidth]{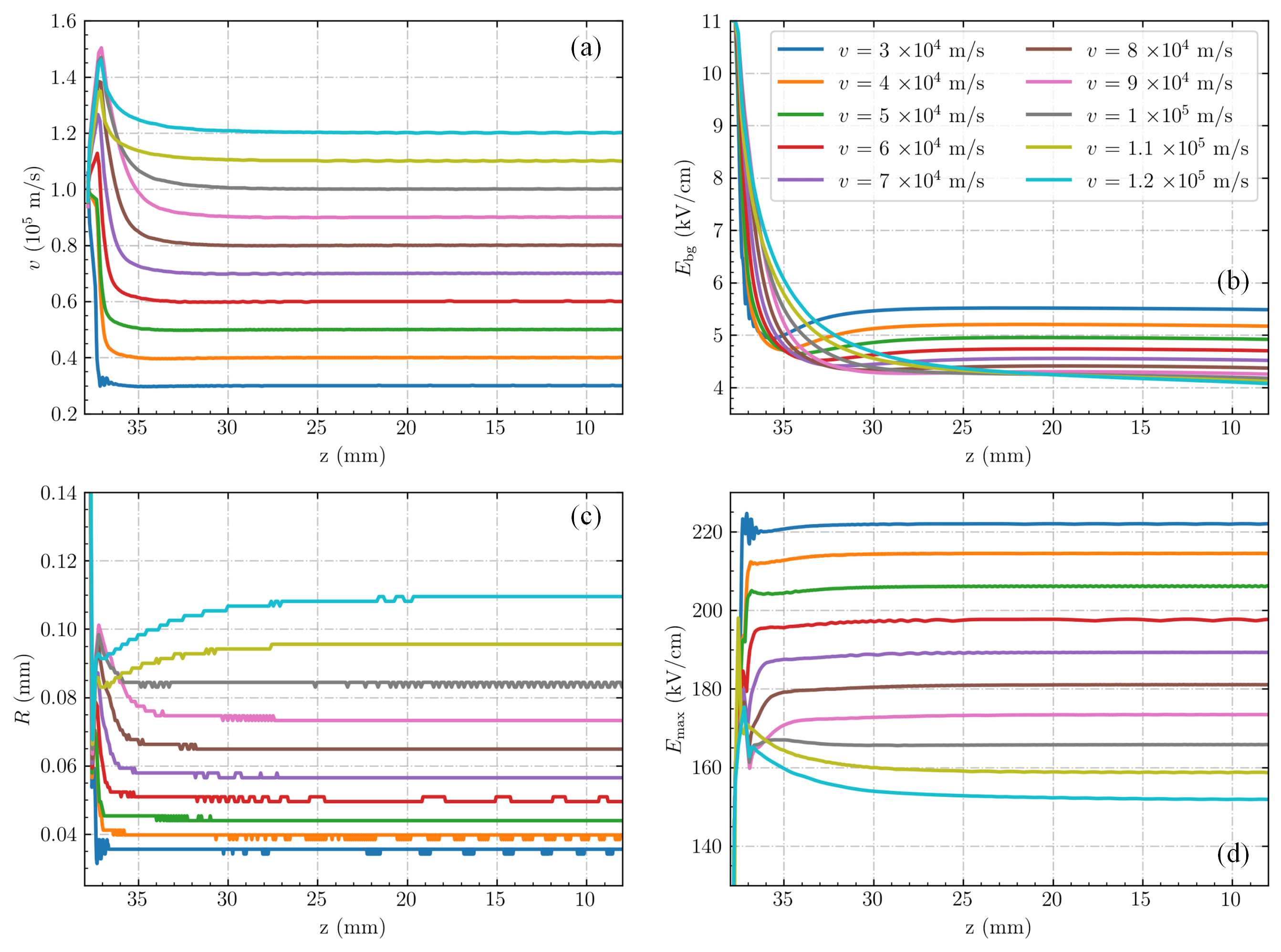}
        \caption{Streamer velocity (a), background electric field (b), streamer radius (c) and the maximal electric field (d) versus the vertical position of the streamer head for streamers in air at velocities from $3\times10^4$\,m/s to $1.2\times10^5$\,m/s.
                We take the location of the maximal electric field as the streamer head position, and the radius as the radial coordinate at which the radial electric field has a maximum. The thinnest (and slowest) streamer has a radius of around 35\,$\mu$m.}
        \label{fig:diffv}
\end{figure*}

\begin{figure*}
        \centering
        \includegraphics[width=1.0\textwidth]{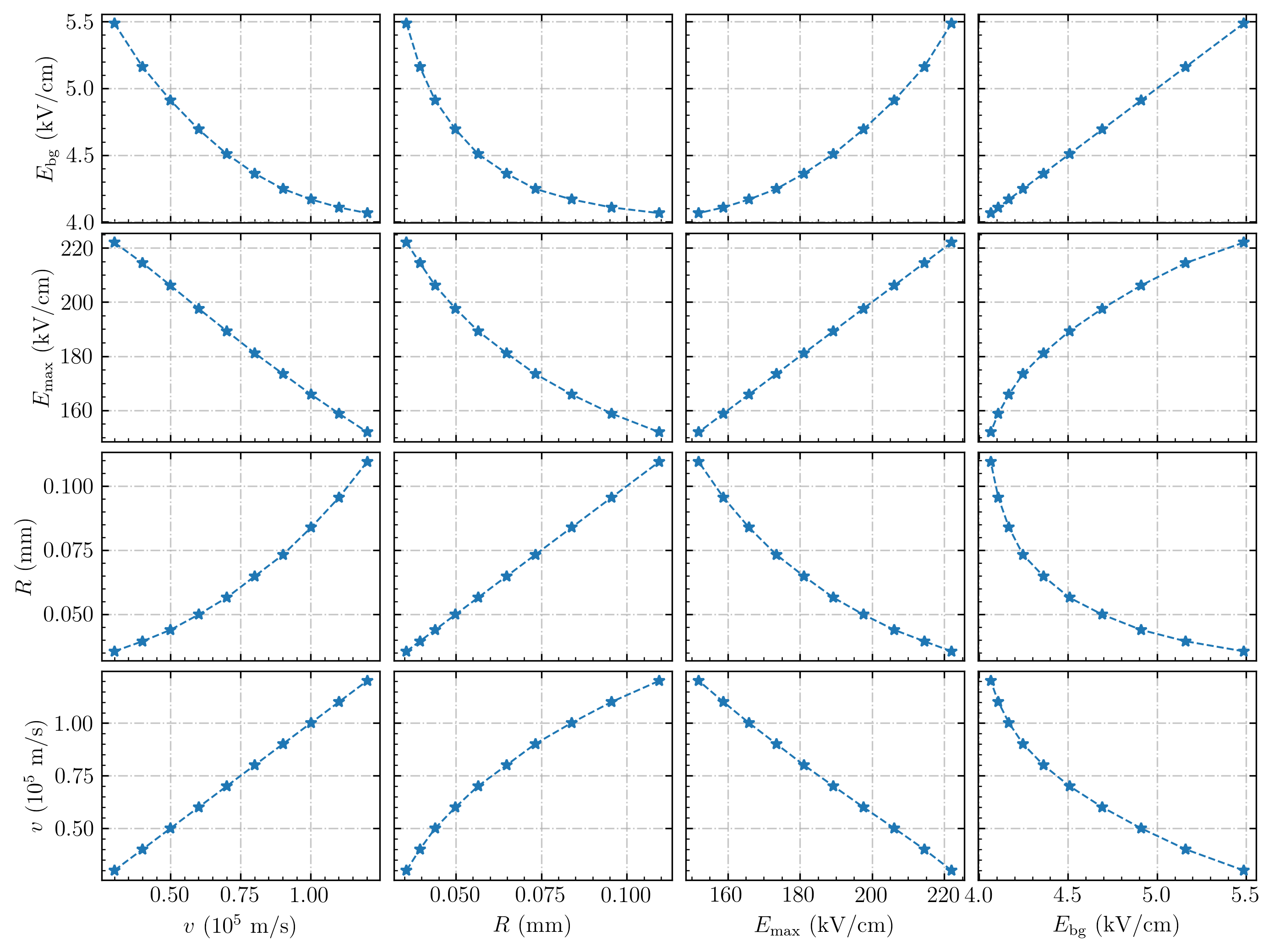}
        \caption{Overview of steady streamer properties. The blue curves show streamer velocity $v$, streamer radius $R$, maximum electric field $E_\mathrm{max}$ and background electric field $E_\mathrm{bg}$. Each star symbol represents a steady propagation state,
          by taking the average over the last 30\,ns of propagation.
          The picture shows each variable as a function of each other variable, with rows sharing the same y-axis and columns the same x-axis.
          }
        \label{fig:diffv_box}
\end{figure*}

We simulate streamers at constant velocities from $3\times10^4$\,m/s to $1.2\times10^5$\,m/s, using the velocity control method described in section~\ref{sec:velocity_control}.
For the two slowest and two fastest streamers, we use $K_p = 5\times10^{13}$\,V/s and $2\times10^{14}$\,V/s, respectively.
	For the streamer at a velocity of $1\times10^5$\,m/s, $K_p = 1\times10^{14}$\,V/s is used.
Figure~\ref{fig:diffv_2d} shows the electric field and the electron density for these streamers when their heads are at z = 16\,mm,
and corresponding on-axis curves are shown in figure~\ref{fig:onaxis_infor}.
Furthermore, figure~\ref{fig:diffv} shows the evolution of the streamer velocity ($v$), radius ($R$), maximal electric field ($E_\mathrm{max}$) and the background electric field ($E_\mathrm{bg}$) versus streamer head position. The streamer radius is here defined as the radial coordinate at which the radial electric field has a maximum.
We remark that there are other definitions of the streamer radius, such as the optical radius and the electrodynamic radius~\cite{pancheshnyi2005a}, which would lead to a different value.
When the streamers reach steady states, $v$, $R$, $E_\mathrm{max}$ and $E_\mathrm{bg}$ all remain constant.
The values corresponding to these steady states are shown versus each other in figure~\ref{fig:diffv_box}.

Faster steady streamers have a larger radius and a lower maximal electric field, but they require a lower background electric field.
For streamer velocities from $3\times10^4$\,m/s to $1.2\times10^5$\,m/s the corresponding background electric fields decrease from 5.5\,kV/cm to 4.1\,kV/cm.
This dependence might at first seem surprising, but can be explained by considering the loss of conductivity in the streamer channel.
Behind the streamer head, electron densities decrease due to attachment and recombination, \xl{and} the electric field relaxes back to the background electric field~\cite{francisco2021d}.
This suggests we can define an effective streamer length as $L_\mathrm{eff} = v \tau$, over which the background electric field is screened,
where $\tau$ is a typical time scale for the loss of conductivity. A faster streamer thus has a longer effective length, as can be seen in figure~\ref{fig:onaxis_infor}. A lower background electric field is therefore sufficient to get a similar amount of electric field enhancement.
That streamers can have a finite conducting length was recently also observed in \cite{francisco2021d}.

With our axisymmetric model, we could not obtain steady streamers faster than $1.2\times10^5$\,m/s due to streamer branching. Another limitation was the limited domain length, due to which the background electric field for the fastest two cases does not become completely constant in figure~\ref{fig:diffv}(b).
Streamers slower than $3\times10^4$\,m/s were also difficult to obtain, because the streamer velocity then becomes comparable to the ion drift velocity at the streamer head, causing the streamers to easily stagnate.
However, the range of steady propagation fields in our simulations agrees well
with the range of experimental stability fields (from 4.14\,kV/cm to 6\,kV/cm)
in~\cite{phelps1971, allen1991, allen1995}.

Our results show that the streamer stability field depends not only on the gas, but also on the streamer properties.
If a faster and wider streamer is able to form, it can propagate in lower background electric fields,
which could explain some of the variation in experimentally determined stability fields in air.
For example, in~\cite{allen1995} streamers were generated from a needle in a plate-plate geometry. It was found that a higher pulse voltage generated faster streamers, which required a lower background electric field to cross the gap.
The minimal steady propagation field in our simulations is about 4.1\,kV/cm.
This value agrees well with the lowest stability fields in
air in previous experimental studies \cite{allen1991, allen1995, seeger2018}.

\subsection{Analysis of steady streamer properties}
\label{sec:dimen_analysis}

Figure \ref{fig:diffv_box} shows streamer velocities, radii, maximal electric fields and background electric fields
corresponding to steady propagation.
Two approximately proportional relations between these variables can be observed. The ratio $E_\mathrm{max}/E_\mathrm{bg}$ is about $40\pm2$, and the ratio $R/v$ is about ($0.95\pm0.2$)\,ns.
Below, we show how these properties can be linked by considering the electric potential difference at the streamer head $\delta \phi$.

First, the effective streamer length can be written as
$L_\mathrm{eff} = v \tau$, where $\tau$ is a characteristic time scale for the
loss of conductivity, see section \ref{sec:diffv}. Just behind the streamer head, the
electric field is almost fully screened, and further behind the head it relaxes back to
the background field. Assuming that the relaxation occurs exponentially, with a characteristic length scale $L_\mathrm{eff}$,
the corresponding potential difference is
\begin{equation}
  \label{eq:phi-channel}
  \delta \phi = v \tau E_\mathrm{bg}.
\end{equation}
Second, the electric field in the vicinity of a streamer head decays approximately quadratically, like that of a charged sphere, with the decay depending on the streamer radius. If one assumes that $E_\mathrm{bg} \ll E_\mathrm{max}$, a simple approximation is given by $E(z) = E_\mathrm{max} (1 + z/R)^{-2}$, with $z = 0$ corresponding to the location of $E_\mathrm{max}$ at the streamer head.
Although this approximation is only justified for $z \lesssim R$, most of the potential drop occurs in this region.
It is therefore not unreasonable to integrate up to $\infty$, giving
\begin{equation}
  \label{eq:phi-emax}
  \delta \phi = \int_0^{\infty} E(z) dz = E_\mathrm{max} R.
\end{equation}
If equations (\ref{eq:phi-channel}) and (\ref{eq:phi-emax}) are combined, the result is
\begin{equation}
  \tau = \frac{E_\mathrm{max} R}{v E_\mathrm{bg}}.
  \label{eq:ts}
\end{equation}

\begin{figure}
  \centering
  \includegraphics[width=0.5\textwidth]{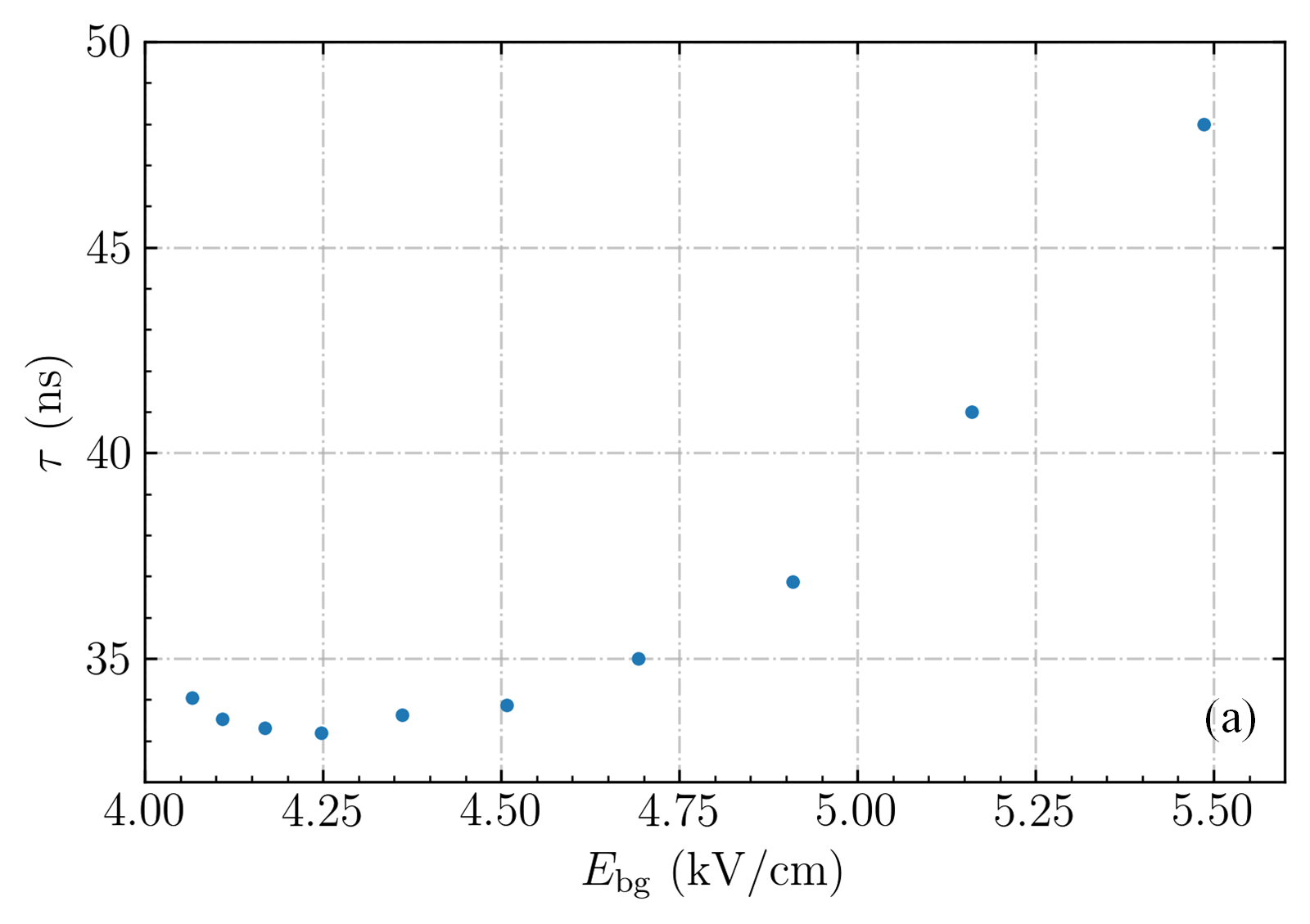}
  \includegraphics[width=0.5\textwidth]{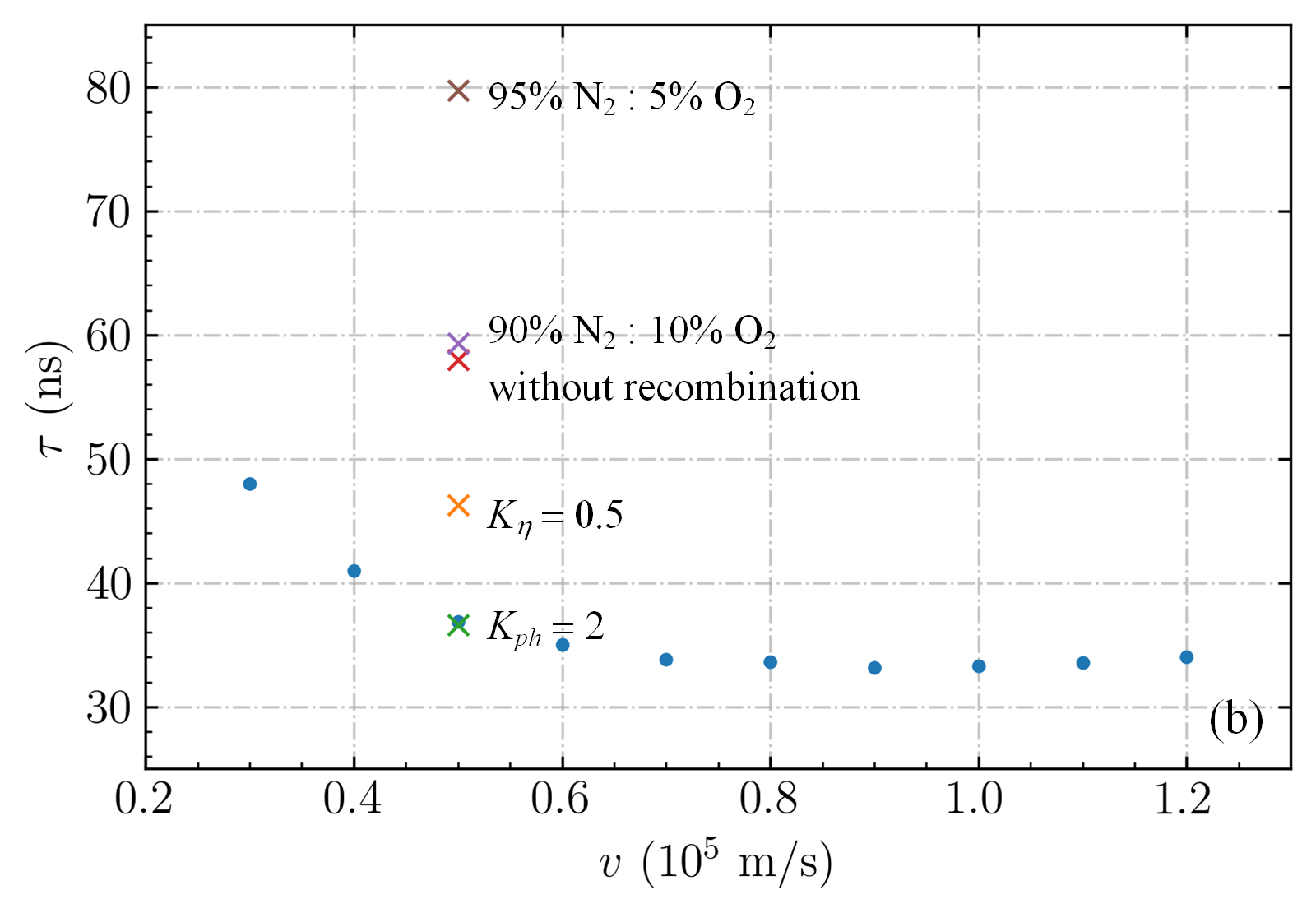}
  \caption{The time scale $\tau$ given by equation (\ref{eq:ts}) versus the background electric field (a) and versus the streamer velocity (b) for the steady streamers in figure~\ref{fig:diffv} and~\ref{fig:difftp_Ebg}.
  }
  \label{fig:timescale}
\end{figure}

Figure~\ref{fig:timescale} shows $\tau$, as defined by equation (\ref{eq:ts}),
versus the background electric field and the streamer velocity. The result lies between 33 and 48\,ns, which
corresponds well with the electron loss time scales due to recombination and
attachment given in~\cite{francisco2021d}. Variation in $\tau$ is to be
expected, because attachment and recombination rates in the streamer channel can
vary, for example due to different electron density and the electric field
profiles. Furthermore, equation (\ref{eq:ts}) was derived based on rather simple
approximations, and does for example not take the degree of ionization produced
by the streamer into account.


For future analysis, we provide additional information on the steady streamers in~\ref{sec:table_steady_streamers}, e.g.,
the maximal electron density, the maximal drift velocity and the maximal ionization rate.

\subsection{Steady streamers in other N$_2$-O$_2$ mixtures}
\label{sec:diffgas}

\begin{figure}
        \centering
        \includegraphics[width=0.5\textwidth]{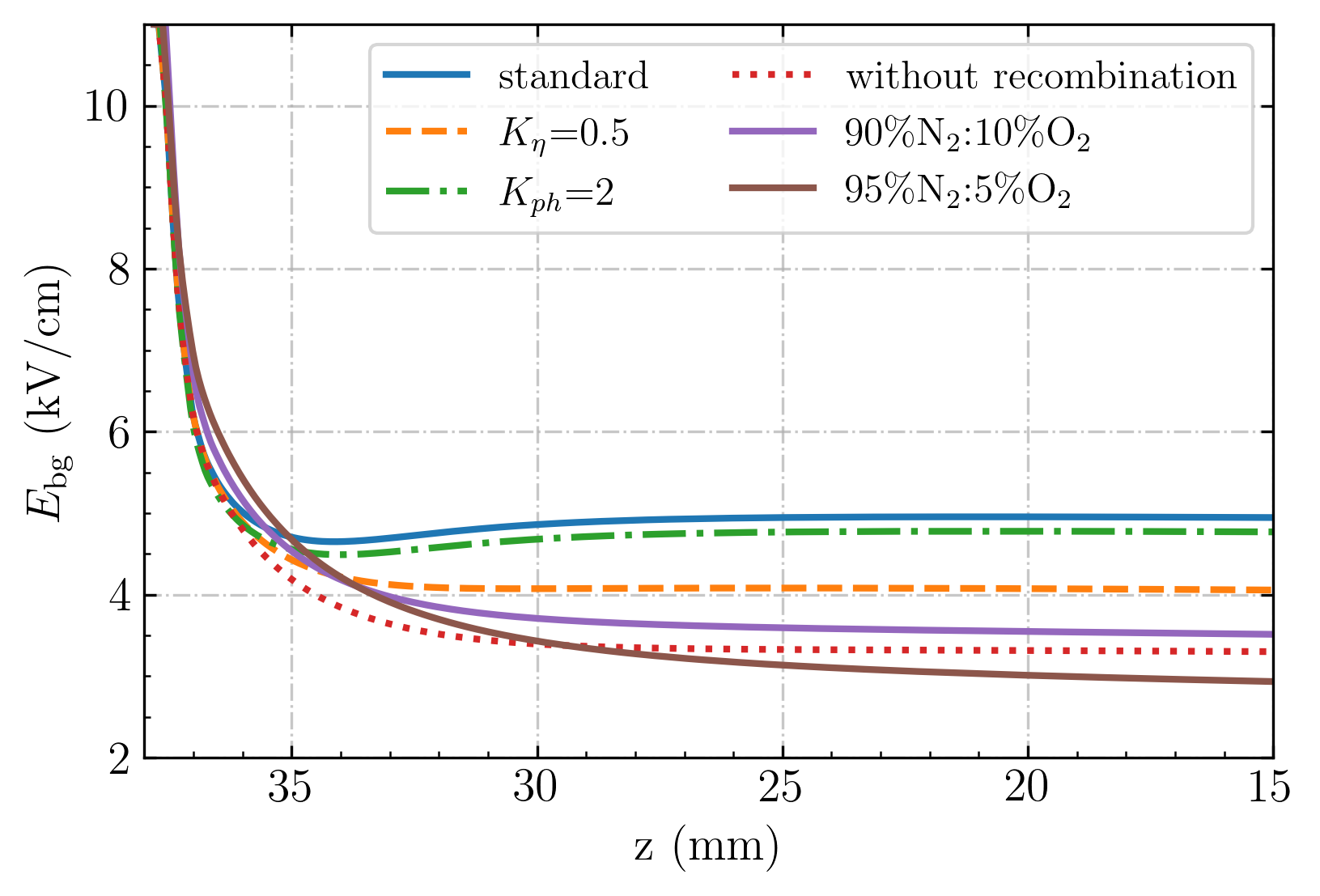}
        \caption{Background electric field versus streamer position for different N$_2$-O$_2$ mixtures and for different transport data. The streamers obtain a constant velocity of $5\times10^4$\,m/s.
                The label ``standard'' indicates the streamer in artificial air (80\%N$_2$:20\%O$_2$).
                The labels ``$K_\eta = 0.5$'' and ``$K_{ph} = 2$'' indicate cases with half the attachment rate and double the amount of photoionization, respectively.}
        \label{fig:difftp_Ebg}
\end{figure}

In this section, we study streamers propagating at $5\times10^4$\,m/s in other N$_2$-O$_2$ mixtures, namely 90\%N$_2$:10\%O$_2$ and 95\%N$_2$:5\%O$_2$, again using the velocity control method but with $T_0$ = 4\,ns. We also consider cases with modified data for air, using either half the attachment rate, double the amount of photoionization or no recombination reactions,
to understand the effect of these processes on the steady propagation mode.
Figure~\ref{fig:difftp_Ebg} shows the background electric field versus streamer position for these cases.
The steady propagation fields for streamers at $5\times10^4$\,m/s are around 4.9\,kV/cm in air, 3.5\,kV/cm in 90\%N$_2$:10\%O$_2$ and 2.9\,kV/cm in 95\%N$_2$:5\%O$_2$.

As shown in figure~\ref{fig:difftp_Ebg}, the effect of doubling the amount of photoionization is rather small. However, both the attachment and the recombination rate have a significant effect on the steady propagation field.
This explains why steady propagation fields are lower with less O$_2$, as attachment and recombination rates are then reduced, see table~\ref{tbl:reaction_table}.
The dominant recombination process in our simulations is between e and O$_4^+$, as O$_4^+$ is one of the main positive ions in the streamer channel~\cite{nijdam2014}.
With less O$_2$, there will also be less O$_4^+$.

The $\times$-symbols in figure~\ref{fig:timescale} show the electron loss time scale $\tau$ given by equation (\ref{eq:ts}) for these steady streamers.
As expected, $\tau$ increases when the attachment rate is halved, when recombination reactions are omitted and when there is less O$_2$.


\section{Investigation of stagnating streamers}
\label{sec:stopping}

Being able to predict whether a streamer can cross a given discharge gap is useful for many applications. In section \ref{sec:invest-steady-stre} we have investigated streamers at constant velocities, which lie at the unstable boundary between acceleration and deceleration. These results help to predict whether a streamer with a certain radius and velocity will accelerate or decelerate, depending on the background field. However, streamers that decelerate might still propagate a significant distance. To predict how far they will go, we need to better understand their deceleration. In this section we therefore simulate decelerating streamers that eventually stagnate.

\subsection{The characteristics of stagnating streamers}
\label{sec:stages}

\begin{figure*}
        \centering
        \includegraphics[width=1.0\textwidth]{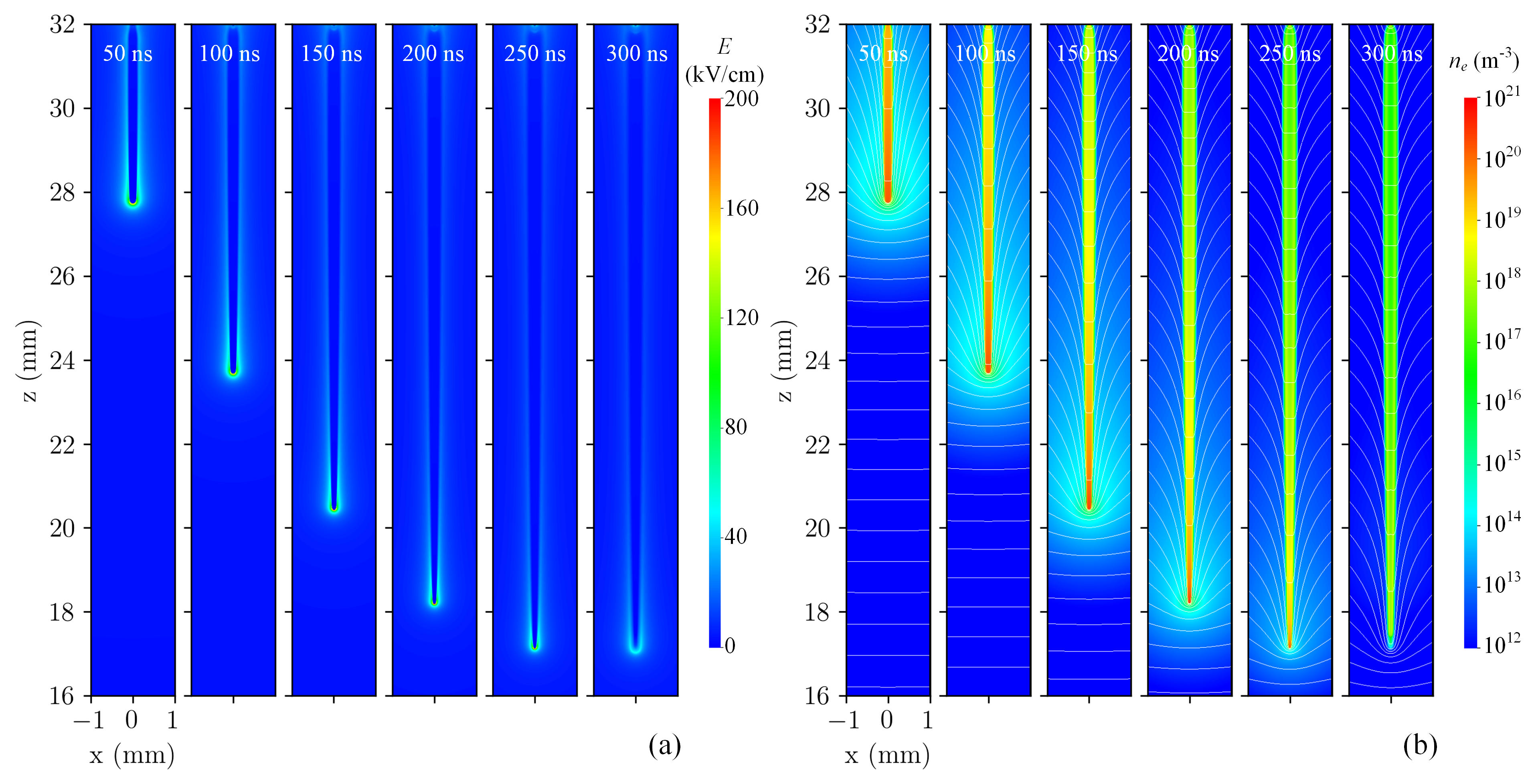}
        \caption{Electric field (a) and electron density (b) of the stagnating positive streamer in air with an applied voltage of 12\,kV between 50\,ns and 300\,ns.
        The white equipotential lines are spaced by 240\,V.}
        \label{fig:stopping_2d}
\end{figure*}

To generate stagnating streamers we use a longer and sharper needle electrode, as described in section~\ref{sec:conditions}. Simulations are performed at constant applied voltages of 11.2, 12 and 12.8\,kV, which correspond to background electric fields of 2.8, 3.0 and 3.2\,kV/cm, respectively. Figure~\ref{fig:stopping_2d} shows the discharge evolution for the 12\,kV case. The streamer decelerates and becomes narrower between 50\,ns and 250\,ns, and it stops after about 250\,ns.
Note that the electric field and electron density at streamer head also decay after 250\,ns, in agreement with~\cite{niknezhad2021}.

\begin{figure*}[th]
	\centering
	\includegraphics[width=0.9\linewidth]{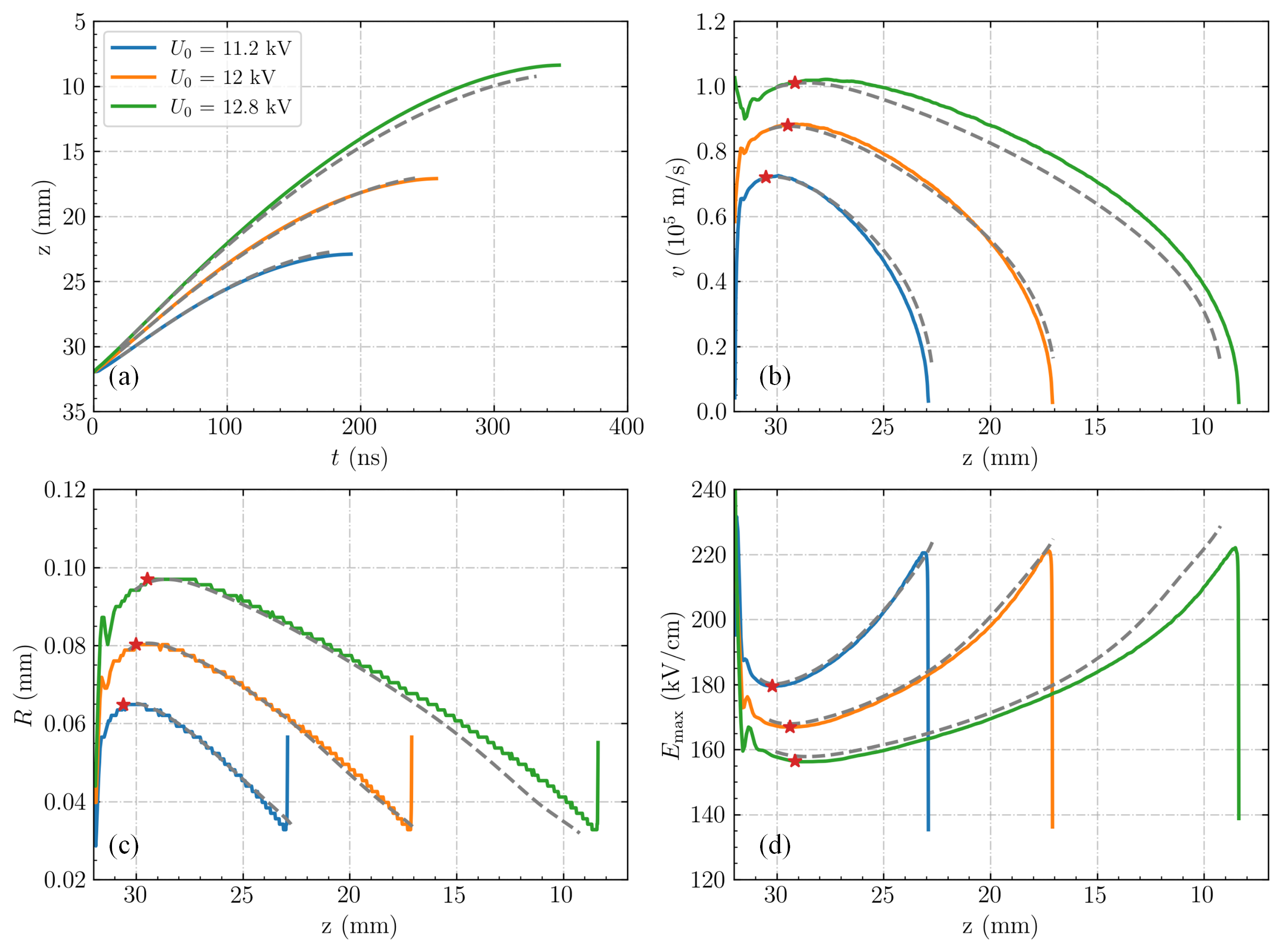}
	\caption{Stagnating streamers in air for applied voltages of 11.2\,kV, 12\,kV and 12.8\,kV. (a) Position versus time,
		(b) streamer velocity, (c) streamer radius and (d) maximum electric field versus streamer position.
		The results of axisymmetric fluid simulations are indicated by solid lines. Solutions of the phenomenological model given by equation (\ref{eq:ode}) are indicated by dashed lines, using $\tau = 38.7$ ns, $\tau_r = 2 \tau$ and $R_\mathrm{min} = 26 \, \mu \mathrm{m}$.
          The red stars indicate the locations where the background field is equal to $E_\mathrm{steady}(v)$ (panel b), $E_\mathrm{steady}(R)$ (panel c) and $E_\mathrm{steady}(E_\mathrm{max})$ (panel d).
          Here $E_\mathrm{steady}(v)$ is the background field corresponding to steady propagation as a function of $v$ (see figure \ref{fig:diffv_box}), and similarly so for $E_\mathrm{steady}(R)$ and $E_\mathrm{steady}(E_\mathrm{max})$.}
	\label{fig:stopping_all}
\end{figure*}

Figure~\ref{fig:stopping_all} shows the evolution of the streamer head position, velocity, radius and maximal electric field for the three stagnating streamers.
As expected, a streamer stops earlier with a lower applied voltage. Several phases can be identified. First there is acceleration in the high field near the electrode, during which $R$ increases and $E_\mathrm{max}$ decreases.
Then there is a transition period, after which the streamer starts to decelerate, with $R$ decreasing and $E_\mathrm{max}$ increasing.
Eventually, the streamer velocity becomes similar to the ion velocity at the streamer head, and the streamer fully stops, as was also observed in~\cite{niknezhad2021}.

With a higher applied voltage, the radius and velocity are larger whereas $E_\mathrm{max}$ is lower. However, the minimal streamer radii are around 32\,$\mu$m for all cases, close to the minimal steady streamer radius in figure \ref{fig:diffv_box}.

\begin{figure}
  \centering
  \includegraphics[width=0.5\textwidth]{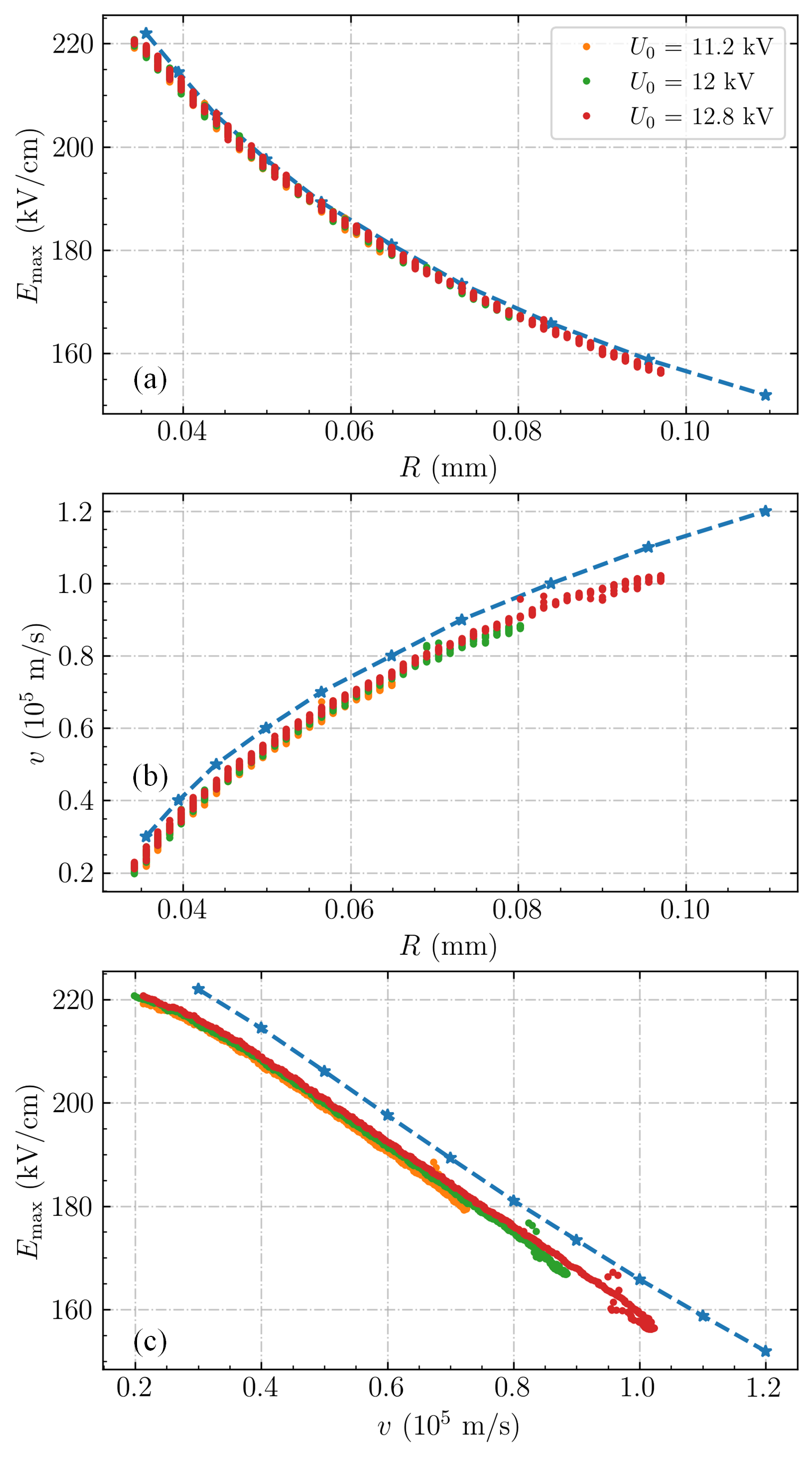}
  \caption{Data for the stagnating streamers shown in
      figure~\ref{fig:stopping_all}, with each data point corresponding to a
      different time. (a) The maximum electric field versus the streamer radius.
      (b) The streamer velocity versus the streamer radius. (c) The maximum
      electric field versus the streamer velocity. Curves for steady streamers \xl{(blue dashed lines)} are shown for comparison. }
  \label{fig:stopping_R_v_Emax}
\end{figure}

In figure~\ref{fig:stopping_R_v_Emax}, we compare temporal $v$, $R$ and $E_\mathrm{max}$ data for the stagnating streamers with data for steady states. 
The relations between $v$, $R$ and $E_\mathrm{max}$ are similar, even though the stagnating streamers develop in lower background fields.
For each of these quantities, the background field corresponding to steady propagation can be obtained from figure \ref{fig:diffv_box}, so that we have functions $E_\mathrm{steady}(v)$, $E_\mathrm{steady}(R)$ and $E_\mathrm{steady}(E_\mathrm{max})$.
In figure~\ref{fig:stopping_all}, we have marked the locations where the actual background field is equal to $E_\mathrm{steady}(v)$, $E_\mathrm{steady}(R)$ and $E_\mathrm{steady}(E_\mathrm{max})$.
Note that at these locations the time derivatives of the respective quantities are approximately zero, as is the case for steady propagation.
In conclusion, the results obtained for steady streamers can help to predict whether a streamer with given properties accelerates or decelerates.

\subsection{A phenomenological model for stagnating streamers}
\label{sec:simple-model}

The deceleration of a streamer depends on multiple properties, e.g., its current velocity, radius and the background electric field.
To quantify this deceleration, we now construct a fitted model for stagnating streamers.
This model is based on their similarity to the steady streamers studied in section \ref{sec:invest-steady-stre}, so we start from equation~(\ref{eq:ts}).
For the data shown in figure \ref{fig:diffv_box}, $E_\mathrm{max}$ can empirically be expressed in terms of the streamer radius as
\begin{equation}
  \label{eq:Emax}
  E_\mathrm{max}(R) = c_0 R^{-1/3} = 7.25 \times 10^5 \left(\frac{1 \, \mathrm{m}}{R}\right)^{1/3} \, \mathrm{V/m},
\end{equation}
with an error below 1\%. Plugging this into equation (\ref{eq:ts}) gives $\tau = c_0 R^{2/3} / (v E_\mathrm{bg})$. Solving for $v$ and $R$ gives the following expressions
\begin{eqnarray}
  \label{eq:steady-v}
  v^* = c_0 R^{2/3}/(E_\mathrm{bg} \tau),\\
  \label{eq:steady-r}
  R^* = \left(E_\mathrm{bg} \tau v/c_0\right)^{3/2}.
\end{eqnarray}
We know that streamers satisfying these equations, for a certain value of $\tau$, keep the same radius and velocity.
A simple coupled differential equation that also satisfies these properties is
\begin{eqnarray}
  \label{eq:ode}
  \partial_t v = \left(v - v^*\right) / \tau_r,\nonumber\\
  \partial_t R = \left(\max[R^*, R_\mathrm{min}] - R \right) / \tau_r,
\end{eqnarray}
with $v^*$ and $R^*$ given as above, $\tau_r$ a relaxation time scale and $R_\mathrm{min}$ a minimal radius. If we fit this model to the stagnating streamer data, rather good agreement is obtained for $\tau = 38.7$ ns, $\tau_r = 2 \tau$ and $R_\mathrm{min} = 26 \, \mu \mathrm{m}$, which all seem reasonable. Solutions for these parameters are shown as dashed lines in figure \ref{fig:stopping_all}. These solutions start at $t = 20$ ns and they take the spatial dependence of the background field into account.
The relatively good agreement suggests that equation (\ref{eq:ode}) can be useful to describe the deceleration of a streamer.





\subsection{Stability field}
\label{sec:stab_field}

We now consider the relation between the streamer length and the concept of a stability field.
The streamer stability field is usually defined as $E_\mathrm{st} = V_0/d$, where $V_0$ is the applied voltage at which a streamer is just able to cross a discharge gap of width $d$~\cite{nijdam2020,allen1991,veldhuizen2002,seeger2018}.
This concept can in principle also be applied to streamers that do not cross the gap, with a length $L_s < d$. A commonly used empirical equation, see e.g.~\cite{seeger2016, bujotzek2015, gallimberti1986}, is
\begin{equation}
\int_{0}^{L_s} (E_\mathrm{bg}(z)-E_\mathrm{st})\, dz = 0,
\label{eq:Ls}
\end{equation}
where $E_\mathrm{bg}$ is the background electric field and the line from $z = 0$ to $z = L_s$ corresponds to the streamer's path.
Equation (\ref{eq:Ls}) can also be written in terms of the background electric potential
$\phi_0(z)$ as $E_\mathrm{st} = (\phi_0(0) - \phi_0(L_s))/L_s$.

For the stagnating streamers at applied voltages of 11.2, 12.0 and 12.8\,kV, the corresponding values of $E_\mathrm{st}$ are 5.10, 4.53 and 4.25\,kV/cm.
These values are in the range of typical observed stability fields.
For a higher applied voltage $E_\mathrm{st}$ is lower, because a faster streamer forms, in agreement with the results of section \ref{sec:invest-steady-stre}.
By using a lower bound for $E_\mathrm{st}$, equation~(\ref{eq:Ls}) can give an upper bound for the streamer length.
If we use $E_\mathrm{st} = 4.1$ kV/cm, as found in section \ref{sec:invest-steady-stre}, the maximal lengths are 16, 21 and 27\,mm for applied voltages of 11.2, 12.0 and 12.8\,kV. For comparison, the actual observed lengths are 9.1, 14.9 and 23.7\,mm, respectively.


The analysis above was based on the background electric field.
In contrast to experimental studies, we can also determine the average electric field inside the streamer channel $\overline{E}_\mathrm{ch}$ in our simulations, including space charge effects.
We measure $\overline{E}_\mathrm{ch}$ as the average field between the electrode and the location where the streamer's electric field has a maximum.
In other words, $\overline{E}_\mathrm{ch} = (\phi_0(0) - \phi(z, t))/L_s$, where $z$ and $t$ are the stagnation location and time, respectively.
This results in average fields of 3.70, 3.63 and 3.68\,kV/cm for applied voltages of 11.2, 12.0 and 12.8\,kV.
These values are significantly lower than the stability fields determined above because they are based on the electrically screened part of the channel.
We can relate $\overline{E}_\mathrm{ch}$ and $E_\mathrm{st}$ by considering the potential difference induced by the streamer head, here denoted as $\delta \phi(z, t) = \phi(z, t) - \phi_0(z)$. It then follows that
\begin{equation}
  \label{eq:headpot}
  \delta \phi(z, t) = L_s \left(E_\mathrm{st} - \overline{E}_\mathrm{ch}\right).
\end{equation}
For applied voltages of 11.2, 12.0 and 12.8\,kV, the corresponding values of $\delta \phi$ are 1.29, 1.41 and 1.37\,kV.
Note that when a streamer crosses a discharge gap, the head potential is zero, so that $E_\mathrm{st} = \overline{E}_\mathrm{ch}$.

In previous work $\overline{E}_\mathrm{ch}$ has been used as a measure of the stability field $E_\mathrm{st}$, even for streamers not crossing the gap~\cite{nijdam2020, briels2008a, morrow1997}. Our results show that $\overline{E}_\mathrm{ch}$ and $E_\mathrm{st}$ can differ significantly. However, according to equation (\ref{eq:headpot}) $\overline{E}_\mathrm{ch}$ and $E_\mathrm{st}$ converge for large streamer length, if ones assume a finite head potential.
Finally, we remark that the value of $\overline{E}_\mathrm{ch} = (\phi_0(0) - \phi(z, t))/L_s$ depends on $z$.
We have here used the location corresponding to the maximal electric field.
Placing $z$ behind the charge layer of the streamer head reduces $\overline{E}_\mathrm{ch}$, whereas placing it further ahead increases $\overline{E}_\mathrm{ch}$, due to the large field around the streamer head.

\section{Conclusions \& Outlook}

\subsection{Conclusions}
\label{sec:conclusion}

We have studied the properties of steady and stagnating positive streamers in air, using an axisymmetric fluid model. \xl{Streamers with constant velocities were obtained by initially} adjusting the applied voltage based on the streamer velocity. Our main findings are listed below.
\begin{itemize}
  \item Positive streamers with constant velocities between $1.2\times10^5$ and $3\times10^4$\,m/s could be obtained in background electric fields from 4.1\,kV/cm to
  5.5\,kV/cm. This range corresponds well with experimentally determined
  stability fields.
  \item The steady streamers are not actually stable, in the sense that a small change in their properties will \xl{eventually} lead to either acceleration or deceleration.
  \item The effective length of a streamer can be described by $v \tau$, where
  $v$ is the streamer velocity and $\tau$ a time scale for the loss of
  conductivity in the streamer channel. A faster streamer has a longer effective
  length, and can therefore propagate in a lower background electric field than
  a slower one.
  \item For the steady streamers, the ratio between radius and
  velocity is about $R/v \sim 0.95\pm0.2$\,ns and the ratio between the maximal
  field at the streamer head and the background field is about
  $E_\mathrm{max}/E_\mathrm{bg} \sim 40\pm2$. However, there is no clear linear
  trend between these variables. To a good approximation,
  $E_\mathrm{max} \propto R^{-1/3}$.
  \item The radius, velocity, maximal electric field and background electric
  field of steady streamers can be related to the conductivity loss
  time scale $\tau$ as $\tau = RE_{max}/vE_\mathrm{bg}$. In air, the obtained values of $\tau$
  range from 33 to 48\,ns.
  \item In N$_2$-O$_2$ mixtures with less O$_2$ than air, steady
  streamers require lower background electric fields, due to reduced attachment
  and recombination rates that result in a longer effective length.
  \item By using a correction factor for the impact ionization source term and by including ion motion, it was possible to simulate stagnating streamers without an unphysical divergence in the electric field.
  \item If a streamer forms near a sharp electrode and then enters a low
  background field, it will first accelerate, then decelerate, and eventually
  stagnate. The transition between acceleration and deceleration occurs close to
  the background electric field corresponding to steady propagation. The
  relationships between $v$, $R$ and $E_\mathrm{max}$ for decelerating streamers are
  similar to those of steady streamers.
  \item A phenomenological model with fitted coefficients was presented to describe the velocity and radius
  of decelerating streamers, based on the properties of steady
  streamers.
  \item For a streamer that has stagnated, the average background electric field
  between the streamer head and tail resembles the empirical stability field.
  The average electric field inside the streamer channel can be significantly
  lower, in particular for relatively short streamers.
\end{itemize}

\subsection{Outlook}

In future work, it would be interesting to include streamer branching and to study the propagation of multiple interacting streamers.
For example, it could be possible that due to repeated branching, streamers in moderately high background field will not continually accelerate, but on average obtain a certain velocity and radius.
Another interesting aspect is how the presence of multiple streamers changes the background field required for their collective propagation, i.e., the stability field.

\section*{Availability of model and data}

The source code and documentation for the model used in this paper are available at \url{gitlab.com/MD-CWI-NL/afivo-streamer} (git commit \texttt{e67bb076}) and at \url{teunissen.net/afivo_streamer}. A snapshot of the code and data is available at \url{https://doi.org/10.5281/zenodo.5873580}.

\section*{Acknowledgments}

X.L. was supported by STW-project 15052 ``Let CO$_2$ Spark'' and the National Natural Science Foundation of China (Grant No. 52077169).


\appendix

\section{Additional information on steady streamers}
\label{sec:table_steady_streamers}

For future analysis, table~\ref{tbl:steady_streamers} provides additional properties of the steady streamers simulated in section \ref{sec:invest-steady-stre}.
All these values are measured at the moments corresponding to figure~\ref{fig:diffv_2d}.
The table contains the following columns:
\begin{itemize}
  \item $E_\mathrm{bg}$ is the steady propagation field
  \item $R$ is the streamer radius, measured as the radius where the radial electric field has a maximum
  \item $E_\mathrm{max}$ is the maximal electric field
  \item $E_\mathrm{min}$ is the minimal electric field in the streamer channel, just behind the streamer head
  \item $v_d(E_\mathrm{max})$ is the electron drift velocity corresponding to $E_\mathrm{max}$
  \item $\max(n_e)$ is the maximum electron density around the streamer head
  \item $n_{\alpha}(E_\mathrm{max})$ is an approximate relation between the maximal
  electric field and the degree of ionization in the streamer
  channel~\cite{ebert1996, babaeva1996}, given by
  \begin{equation}
    \label{eq:alphaint}
    n_{\alpha}(E_\mathrm{max}) = \frac{\varepsilon_0}{e} \int_{0}^{E_\mathrm{max}} \alpha(E) dE,
  \end{equation}
  where $\alpha$ is the ionization coefficient. For positive streamers,
  $n_{\alpha}(E_\mathrm{max})$ has been observed to be about half the degree of
  ionization in the channel~\cite{nijdam2020}.
  \item $S(E_\mathrm{max})$ is the ionization rate corresponding to $E_\mathrm{max}$
  \item $\delta\phi$ is the potential difference at the streamer head, defined as $\phi(z, t) - \phi_0(z)$, with $z$ the location corresponding to $E_\mathrm{max}$.
\end{itemize}


\begin{table*}
  \centering
  \caption{Properties of the steady streamers simulated in section
    \ref{sec:invest-steady-stre}, see \ref{sec:table_steady_streamers} for a
    description of the columns.}
  \begin{tabular}{c c c c c c c c c c}
    \hline
    \makecell[c]{velocity \\(m/s)}
    & \makecell[c]{$E_\mathrm{bg}$ \\(kV/cm)}
    & \makecell[c]{$R$ \\($\mu$m)}
    & \makecell[c]{$E_\mathrm{max}$ \\(kV/cm)}
    & \makecell[c]{$E_\mathrm{min}$ \\(kV/cm)}
    & \makecell[c]{$v_d(E_\mathrm{max})$ \\(m/s)}
    & \makecell[c]{$\max(n_e)$\\ (m$^{-3})$}
    & \makecell[c]{$n_{\alpha}(E_\mathrm{max})$ \\ (m$^{-3})$}
    & \makecell[c]{$S(E_\mathrm{max})$ \\ (s$^{-1}$)}
    & \makecell[c]{$\delta\phi$\\ (kV)} \\
    \hline
    $3\times10^4$   & 5.48 & 36  & 222 & 0.42 & $6.1 \times 10^5$ & 3.13$\times$10$^{20}$ & 1.69$\times 10^{20}$ & 2.24$\times$10$^{11}$ & 1.51 \\
    $4\times10^4$   & 5.16 & 39  & 214 & 0.50 & $5.9 \times 10^5$ & 2.95$\times$10$^{20}$ & 1.54$\times 10^{20}$ & 2.10$\times$10$^{11}$ & 1.60 \\
    $5\times10^4$   & 4.91 & 44  & 206 & 0.56 & $5.8 \times 10^5$ & 2.70$\times$10$^{20}$ & 1.39$\times 10^{20}$ & 1.93$\times$10$^{11}$ & 1.70 \\
    $6\times10^4$   & 4.69 & 50  & 198 & 0.66 & $5.6 \times 10^5$ & 2.43$\times$10$^{20}$ & 1.23$\times 10^{20}$ & 1.77$\times$10$^{11}$ & 1.82 \\
    $7\times10^4$   & 4.51 & 57  & 189 & 0.74 & $5.4 \times 10^5$ & 2.16$\times$10$^{20}$ & 1.09$\times 10^{20}$ & 1.61$\times$10$^{11}$ & 1.96 \\
    $8\times10^4$   & 4.36 & 65  & 181 & 0.82 & $5.3 \times 10^5$ & 1.91$\times$10$^{20}$ & 9.61$\times 10^{19}$ & 1.46$\times$10$^{11}$ & 2.11 \\
    $9\times10^4$   & 4.25 & 73  & 173 & 0.94 & $5.1 \times 10^5$ & 1.68$\times$10$^{20}$ & 8.48$\times 10^{19}$ & 1.33$\times$10$^{11}$ & 2.28 \\
    $1\times10^5$   & 4.16 & 84  & 166 & 1.0  & $5.0 \times 10^5$ & 1.46$\times$10$^{20}$ & 7.42$\times 10^{19}$ & 1.20$\times$10$^{11}$ & 2.61 \\
    $1.1\times10^5$ & 4.11 & 96  & 159 & 1.08 & $4.8 \times 10^5$ & 1.28$\times$10$^{20}$ & 6.51$\times 10^{19}$ & 1.09$\times$10$^{11}$ & 2.64 \\
    $1.2\times10^5$ & 4.07 & 110 & 152 & 1.17 & $4.7 \times 10^5$ & 1.11$\times$10$^{20}$ & 5.68$\times 10^{19}$ & 9.79$\times$10$^{10}$ & 2.83 \\
    \hline
  \end{tabular}
  \label{tbl:steady_streamers}
\end{table*}

\section*{References}

\bibliography{references_zotero}

\end{document}